\documentstyle[prl,tighten,aps,epsf,multicol,palatino]{revtex}
\begin{document}

\draft

\title{Catalysis of entanglement manipulation for mixed states}
\author{Jens Eisert and Martin Wilkens}
\address{Institut f{\"u}r Physik, Universit{\"a}t Potsdam, 14469 Potsdam,
Germany}
\date{\today}
\maketitle

\begin{abstract}
We consider entanglement-assisted remote quantum 
state manipulation of bi-partite mixed states.
Several aspects are addressed:
we present a class of mixed states of
rank two that can be transformed into another class of 
mixed states under entanglement-assisted local
operations with classical communication,
but for which such a transformation
is impossible without assistance.
Furthermore, we demonstrate enhancement 
of the efficiency of purification protocols
with the help of entanglement-assisted operations. 
Finally, transformations from one mixed 
state to mixed
target states which are sufficiently close
to the source state are contrasted to
similar transformations in the pure-state case.
\end{abstract}


\bigskip\medskip

\begin{multicols}{2}
\narrowtext


\section{Introduction}
Entanglement between spatially separated quantum systems
has important implications on fundamental issues of quantum
mechanics and forms the basis for most of 
the practical applications of quantum information
theory \cite{Contemp,Practical}. In many of
these applications two or more parties have direct
access to only
parts of a composite quantum system, but may communicate
by classical means and may thereby 
coordinate their actions. 
In the light of recent 
progress in quantum information
theory entanglement is often viewed as the essential resource 
for processing and transmitting quantum information.

As has been demonstrated in Ref.\ \cite{Jonathan},
entanglement is indeed an intriguing kind of resource:
the mere presence of entanglement can be
an advantage when the task it to 
transform an initial state into a certain
final state 
with the use of local quantum operations and 
classical communication (LOCC).
There are indeed target states which cannot be 
reached by LOCC starting from a particular
initial state, but which can be reached
with the assistance of a
distributed pair of auxiliary quantum systems in a 
particular known state, 
%
%
even though these auxiliary quantum systems
are left in {\it exactly the same state}\/.
Such transformations are called entanglement-assisted
LOCC operations (ELOCC).

This phenomenon is quite remarkable as the entanglement
which serves as a ''catalyst'' for the otherwise
forbidden ''reaction'' is not consumed. 
The basis of the example given in Ref.\ \cite{Jonathan}
is a criterion 
presented in Ref. \cite{Nielsen}: 
A joint pure state corresponding to 
$|\psi\rangle$ can be transformed into another 
$|\phi\rangle$ with the use of LOCC if
and only if the set of ordered Schmidt coefficients 
characterizing
the initial state is majorized \cite{Majo} by the set
of  ordered Schmidt coefficients of the final state.
Curiously, it is the strange class of 
ELOCC operations
that adds a new flavor to the initial question 
raised in  Ref.\ \cite{Nielsen},
``What tasks may be accomplished using a given 
physical resource?'' The class of ELOCC operations
is in fact more powerful than LOCC even without 
a concomitant consumption of the
physical resource entanglement \cite{Jonathan,NewDaniel}.

In practical applications one
would expect to always deal with entangled mixed states 
rather than with pure states.
Unfortunately, such a convenient tool as the majorization
criterion is missing in the mixed-state case, and
the question whether a particular entanglement transformation
from one mixed state into another mixed state is possible seems 
to be much more involved \cite{After}. 
In mixed quantum mechanical states 
both classical correlations and
intrinsic quantum correlations may be present, which
makes the structure of mixed-state entanglement a more
complex matter. A different aspect of the same problem is the
well known fact that a representation of a mixed state in terms
of pure states is not uniquely defined, and
it is essentially 
this ambiguity that prohibits a straightforward
application of the majorization criterion.

In this letter we demonstrate that even for mixed states the
set of tasks that can be accomplished with 
entanglement-assisted local operations is strictly larger than the set
of tasks which may be performed with mere LOCC.
This fact is not obvious a priori,
bearing in mind that e.g.\ 
pure states and mixed states behave very differently
as far as purification is concerned \cite{Kent}. 
The problem of catalysis of 
entanglement manipulation for mixed states
will be approached as follows:
(i) We give a class of mixed states of rank two
that can be transformed into representants of
another class of mixed states with ELOCC but
not with LOCC, (ii) we show that
there are cases for which 
the proportion of a certain pure state in a mixture
can be increased more efficiently with ELOCC
operations than with sole LOCC, 
(iii) purification schemes are investigated for
a practically important class of mixed states, and (iv)
``small transformations'' in the interior of the state
space are compared with
similar entanglement manipulations in the pure-state
case.

\section{Definitions}
Let $\sigma$ and $\rho$ be states taken from
the state space ${\cal S}({\cal H})$ over ${\cal H}$,
where ${\cal H}={\cal H}_A\otimes {\cal H}_B$ is the 
Hilbert space associated with a bipartite quantum system
consisting of parts $A$ and $B$.
We write in the following $\sigma\rightarrow\rho$ under
LOCC if $\sigma$ can be transformed into $\rho$ by applying local
transformations and classical communication \cite{Nielsen}.
A pair of states $\rho,\sigma$ is called
incommensurate if both $\sigma\not\rightarrow\rho$
and $\rho\not\rightarrow\sigma$ under LOCC.
For {\it pure}\/ states $\sigma$ and $\rho$ the
(necessary and sufficient) majorization
criterion 
for $\sigma\rightarrow\rho$
under LOCC reads as \cite{Nielsen}
\begin{equation}
	\sum_{i=1}^k \alpha_i\leq \sum_{i=1}^k \beta_i
	\,\,\,\,\,\,\,\,\,
	$ for all $
	k=1,...,N-1,\label{maj}
\end{equation}
$N=\dim[{\cal H}_A]=\dim[{\cal H}_B] $, where $\alpha_1$, ..., $\alpha_N$
and $\beta_1$, ..., $\beta_N$ with
$1\geq\alpha_1\geq ...\geq\alpha_N\geq0$ and
$1\geq\beta_1\geq...\geq \beta_N\geq0$ are
the eigenvalues of ${\rm tr}_A [\sigma]$
and ${\rm tr}_A [\rho]$, respectively. Such a list
is also referred to as an ordered list.
The content of the conditions stated in Eq.\ (\ref{maj})
is in the following abbreviated as 
${\rm tr}_A [\sigma]\prec {\rm tr}_A [\rho]$,
with the majorization relation $\prec$ \cite{Majo}.
As for LOCC operations we use the notation 
$\sigma\rightarrow\rho$ under ELOCC,
if 
\begin{equation}
	\sigma\otimes\omega\rightarrow\rho\otimes\omega 
\end{equation}
for an appropriately chosen catalyst state $\omega$  
\cite{Jonathan}.
This state $\omega$ is an entangled state of 
another bi-partite quantum system. Note that in the 
course of the transformation this state 
remains fully unchanged.


\section{Mixed-state catalysis of entanglement manipulation}
The first result concerns the existence of
incommensurate genuinely mixed states such
that with the use of some appropriately chosen
catalyst state, the initial state can be converted into 
the final state while fully retaining the
catalyst state.  That is, there exist 
mixed 
states $\sigma,\rho\in{\cal S}({\cal H})$ 
such that
        $\sigma\rightarrow \rho$ under ELOCC
but not
        $\sigma\rightarrow \rho$ under LOCC.
``Genuinely'' mixed means here that 
the projections appearing in the spectral
decomposition of the initial state
cannot be locally distinguished. If this
were possible the 
initial state would essentially be pure.

To see that mixed-state catalysis is possible we 
construct a class of states which exhibits
this phenomenon. For this class of 
states the statement that 
$\sigma\rightarrow\rho$ under ELOCC
follows immediately from 
the theorem presented in Ref.\ \cite{Nielsen}.
To prove that such a transformation is
impossible under LOCC, the following 
Lemma is useful.\\

\noindent{\bf Lemma 1. --}  Let $\sigma$ and $\rho$ 
be mixed states of rank two of the form 
\begin{mathletters}
\begin{eqnarray}
\sigma&=&\lambda |\psi\rangle\langle\psi|
+(1-\lambda)|\eta\rangle\langle\eta|,\\
\rho&=&\mu 
|\phi\rangle\langle\phi|
+(1-\mu)|\eta\rangle\langle\eta|,\label{class}
\end{eqnarray}
where $\mu=\lambda \,{\rm tr}[\chi]$, 
\end{mathletters}
\begin{equation}
	\chi=\Pi|\psi\rangle\langle\psi|\Pi,
\end{equation} 
and
$\Pi=1-|\eta\rangle\langle\eta|$. 
$|\psi\rangle\langle\psi|$ and $|\phi\rangle\langle\phi|$ are 
entangled pure states, while
$|\eta\rangle\langle\eta|$ is a  pure product state. 
Furthermore, 
$|\langle\eta|\phi\rangle|^2=0$.
Then $\sigma\rightarrow\rho$ under
LOCC implies that 
\begin{equation}
	\frac{{\rm tr}_A [\chi]}{{\rm tr} [\chi]}
	 \prec
	{\rm tr}_A [|\phi\rangle\langle\phi|].
\end{equation}

{\it Proof:} 
Assume that $\sigma\rightarrow\rho$ under
LOCC.
The set of LOCC operations is included in the set
of separable operations \cite{Plenio,Rains}, that is,
completely positive and trace-preserving maps
that can be written in the form 
$\sigma\longmapsto
\sum_i (A_i\otimes B_i)\sigma (A_i\otimes B_i)^\dagger$
with Kraus-operators $A_i$, $B_i$, $i=1,2,...$, acting
in ${\cal H}_A$ and ${\cal H}_B$, respectively, 
where the trace-preserving property manifests as
$\sum_i A_i^\dagger A_i=1$, $\sum_i B_i^\dagger B_i=1$.
For each $i$ the image of $\sigma$ must be element
in the range of $\rho$,
\begin{equation} 
(A_i\otimes B_i)\sigma (A_i\otimes B_i)^\dagger\in {\rm range}(\rho).
\end{equation} 
Since there is only a single product
vector included in the range of $\rho$ (which 
then amounts to a best separable approximation in the
sense of \cite{Lewenstein}), 
the state
$|\psi\rangle\langle\psi|$ must be mapped on
$\nu |\phi\rangle\langle\phi|+(1-\nu)
|\eta\rangle\langle\eta|$, where $\nu=\mu/\lambda$.
$\Pi (A_i\otimes B_i) |\psi\rangle=
 \Pi (A_i\otimes B_i) \Pi |\psi\rangle$
for all $i$, and hence,
\begin{eqnarray}
\nu&=&{\rm tr}\bigl[
\Pi\sum_i (A_i\otimes B_i) |\psi\rangle\langle\psi|(A_i\otimes B_i)^\dagger
\Pi\bigr]\nonumber\\
&=&{\rm tr}\bigl[ \sum_i \Pi (A_i\otimes B_i) \chi (A_i\otimes B_i)^\dagger
\Pi\bigr]\leq{\rm tr}[\chi].
\end{eqnarray}
As ${\rm tr}[\chi]=\nu$, it follows that
$\chi/{\rm tr}[\chi]\longrightarrow |\phi\rangle\langle\phi|$
under LOCC, which in turn implies
by the theorem in Ref.\ \cite{Nielsen}
that ${\rm tr}_A [\chi]/{\rm tr} [\chi]
	 \prec
	{\rm tr}_A [|\phi\rangle\langle\phi|]$.
$\hfill{\Box}$

The following one-parameter classes of states of rank two
provide an  example of catalysis for mixed states.
Take
${\cal H}={\cal H}_A\otimes {\cal H}_B$ with
${\cal H}_A, {\cal H}_B=
\text{span}\{|1\rangle, ...,|5\rangle \}$
and let 
\begin{mathletters}
\begin{eqnarray}\label{example}
\sigma&=&\lambda|\psi\rangle\langle\psi|+(1-\lambda)|55\rangle\langle55|,\\
\rho&=&\mu |\phi\rangle\langle\phi|
+(1-\mu)|55\rangle\langle55|,
\end{eqnarray}
with $\mu=0.95\;\lambda$ and 
\end{mathletters}
\begin{mathletters}
\begin{eqnarray}
	|\psi\rangle&=&
	\sqrt{0.38}|11\rangle+
	\sqrt{0.38}|22\rangle+
	\sqrt{0.095}|33\rangle\nonumber \\
	&+&
	\sqrt{0.095}|44\rangle+
	\sqrt{0.05} |55\rangle,\label{psi}\\
	|\phi\rangle&=&
	\sqrt{0.5}|11\rangle+
	\sqrt{0.25}|22\rangle+
	\sqrt{0.25}|33\rangle.\label{phi}
\end{eqnarray}
These states are
clearly included in the sets of states considered
in Lemma 1. 
\end{mathletters}
Moreover, the initial state $\sigma$ is
genuinely mixed. 

From Lemma 1 it follows that 
$\sigma\not\rightarrow\rho$ under LOCC
for all values of $\lambda\in(0,1]$, 
as $\chi/{\rm tr}[\chi]=|\varphi\rangle\langle\varphi|$,
where 
\begin{equation}\label{tild}
|\varphi\rangle=\sqrt{0.4}|11\rangle+
\sqrt{0.4}|22\rangle+
\sqrt{0.1}|33\rangle+
\sqrt{0.1}|44\rangle
\end{equation}
as in Ref. \cite{Jonathan}. Hence, 
\begin{equation}
	\frac{{\rm tr}_A [\chi]}{{\rm tr} [\chi]}
	 \not\prec
	{\rm tr}_A [|\phi\rangle\langle\phi|],
\end{equation}
and therefore, $\sigma\not\rightarrow\rho$
under LOCC.
However, it can be shown that
$\sigma\rightarrow\rho$ under ELOCC. One 
may perform a local projective von-Neumann measurement
in system $A$
associated with Kraus operators 
$A_1=\sum_{i=1}^4 |ii\rangle\langle ii|$ and
$A_2=|55\rangle\langle55|$
satisfying 
$A_1^\dagger A_1+
A_2^\dagger A_2=1$
(compare also 
\cite{Amnesia}). 
If one gets the outcome
corresponding to $A_2$, no further 
operations are applied. 
In the other case the final state is the pure state 
$
|\varphi\rangle\langle\varphi|$ given by
Eq.\ (\ref{tild}). 
As in Ref.\ \cite{Jonathan} this state can be
transformed into $|\phi\rangle\langle\phi|$
with the help of the catalyst state $\omega=(\sqrt{0.4}|66\rangle+
\sqrt{0.6}|77\rangle)(\sqrt{0.4}\langle66|+
\sqrt{0.6}\langle77|)$ \cite{HilbertRemark},
since 
\begin{equation}
{\rm tr}_A [|\varphi\rangle\langle\varphi|
\otimes \omega]\prec 
{\rm tr}_A [|\phi\rangle\langle\phi|
\otimes \omega].
\end{equation} 
Finally, 
the classical information about the outcomes is discarded
in order to achieve $\rho$. 
Hence, it turns out that $\sigma\rightarrow\rho$ under ELOCC but 
$\sigma\not\rightarrow\rho$ under LOCC.

\section{Increasing the proportion of a pure state in a mixture}

The possibility of catalysis of entanglement manipulations
has an implication 
on the efficiency of attempts to 
increase the quota of some entangled 
state $|\xi\rangle\langle\xi|$ in a mixed state $\sigma$
by applying a trace-preserving operation. 
Indeed, such protocols can be more efficient
when 
employing ELOCC rather than exclusively using LOCC.
More precisely, there are 
(genuinely) mixed states $\sigma$ and
pure states $|\xi\rangle\langle\xi |$ with the property that 
the maximal average attainable value of the fidelity 
under ELOCC
\begin{equation}
	F_{\rm ELOCC}(\sigma,|\xi\rangle\langle\xi|)= 
	\sup_{\rho\in{\cal S}_{\rm ELOCC}^\sigma}
	\langle \xi|\rho|\xi\rangle
\end{equation}
is strictly larger than the 
maximal attainable fidelity under LOCC,
\begin{equation}
F_{\rm LOCC}(\sigma,|\xi\rangle\langle\xi|)
= \sup_{\rho\in{\cal S}_{\rm LOCC}^\sigma}\langle \xi|\rho|\xi\rangle.
\end{equation}
Here,  ${\cal S}_{\rm LOCC}^\sigma$ 
and ${\cal S}_{\rm ELOCC}^\sigma$
are the sets of states that can be reached by applying LOCC and
ELOCC, respectively, on an initial state $\sigma$. 

This statement can be proven by considering
an initial state $\sigma$ 
of the form specified in
Eq.\ (\ref{example}) with
\begin{eqnarray}
	|\psi\rangle&=&
	\varepsilon(\sqrt{0.4}|11\rangle+
	\sqrt{0.4}|22\rangle+
	\sqrt{0.1}|33\rangle+
	\sqrt{0.1}|44\rangle)\nonumber\\
	&+&
	\sqrt{1-\varepsilon^2}|55\rangle,\label{fidexample}
\end{eqnarray}
and one may choose $|\xi\rangle=|\phi\rangle$ as in Eq.\ (\ref{phi}).  
Clearly
\begin{eqnarray}
F_{\rm LOCC}(\sigma,|\phi\rangle\langle\phi|)&\leq&
(1-\lambda)
F_{\rm LOCC}(|55\rangle\langle55|,|\phi\rangle\langle\phi|)\nonumber\\
&+&
\lambda
F_{\rm LOCC}(|\psi\rangle\langle\psi|,|\phi\rangle\langle\phi|),
\end{eqnarray}
as the components of the initial state $\sigma$ are
not locally distinguishable, and since the achievable
fidelity can be no better than the sum of 
both best possible fidelities of each contribution.
Under LOCC
all separable states are accessible starting from
$|55\rangle\langle55|$.
The (not necessarily pure) 
separable state closest to $|\phi\rangle\langle\phi|$
with respect to the fidelity is given by $|11\rangle\langle11|$,
and therefore, 
$F_{\rm LOCC}(|55\rangle\langle55|, |\phi\rangle\langle\phi|)=
1/2$.
Finally, from $F_{\rm ELOCC}(\sigma,|\phi\rangle\langle\phi|)\geq
\lambda\varepsilon^2+(1-\lambda\varepsilon^2)/2$ it follows that
\begin{equation}
F_{\rm ELOCC}(\sigma,|\phi\rangle\langle\phi|)>
F_{\rm LOCC}(\sigma,|\phi\rangle\langle\phi|)
\end{equation}
certainly
holds for all $\varepsilon\in(\tilde\varepsilon,1]$, where
\begin{equation}
\tilde\varepsilon
=(2F_{\rm LOCC}(|\psi\rangle\langle\psi|,
|\phi\rangle\langle\phi|)-1 )^{1/2},
\end{equation}
independent of $\lambda\in(0,1)$, and for all $\varepsilon<1$ the
initial state is also genuinely mixed.

\section{Purification procedures} 
The previous two results unambiguously indicate that
the class of ELOCC operations is more powerful than
LOCC operations 
not only on the subset of
the boundary of ${\cal S}({\cal H})$ comprising 
the pure states, but 
also in the interior of the set 
${\cal S}({\cal H})$. Albeit this facts suggests 
that the use of supplementary catalyst states opens up
possibilities to enhance purification procedures,
ELOCC do not necessarily imply an improved efficiency
in practically motivated
problems.
Consider the class of states studied in Ref.\ \cite{Kent}
\begin{equation}	
	\sigma=\lambda |\psi\rangle\langle\psi|+(1-\lambda)\zeta
\end{equation}
with the property that there exists a 
$\lambda_0\in(0,1)$ such that $\sigma$
is a separable state and that every state with
a larger weight of $|\psi\rangle\langle\psi|$ is
entangled.  Furthermore, it is assumed that
$\langle\psi|\zeta|\psi\rangle=0$. 
This class of states 
includes the class of states consisting
of a mixture of some pure state and the complete mixture
in the corresponding state space, which is of salient
importance in practical applications.
In Ref.\ \cite{Kent} is has been shown that
$\langle\psi|\rho|\psi\rangle\leq \langle\psi|\sigma|\psi\rangle$
for all states $\rho$ that can be reached from $\sigma$
with {\it any probability}\/ $p>0$
(that is, not necessarily $\sigma\rightarrow\rho$ under LOCC holds),
implying that for this class of states the proportion of
$|\psi\rangle\langle\psi|$ can not even be increased
with non-trace-preserving operations \cite{MachineRemark}. 
This is
also true for ELOCC operations.

Let $\sigma\in{\cal S}({\cal H})$ be such a state,
and let $\omega\in {\cal S}(\tilde{\cal H})=
{\cal S}(\tilde{\cal H}_A\otimes\tilde{\cal H}_B )$
be an appropriate catalyst state. The above
transformation then amounts to a map
\begin{equation}\label{trans}
	\sigma\otimes \omega\longmapsto
	\rho\otimes\omega=
	\frac{\sum_i
	(A_i\otimes B_i)(\sigma\otimes \omega)(A_i\otimes B_i)^\dagger}
	{{\rm tr} [\sum_i
	(A_i\otimes B_i)(\sigma\otimes \omega)(A_i\otimes B_i)^\dagger]},
	\end{equation}
where $A_i$ and $B_i$ satisfying $\sum_i A_i^\dagger A_i\leq 1$ and
$\sum_i B_i^\dagger B_i\leq 1$
act only in ${\cal H}_A\otimes \tilde{\cal H}_A$ and 
${\cal H}_B\otimes \tilde{\cal H}_B$, respectively. 
The quantity of
interest is now the fidelity $F=\langle\psi|\rho|\psi\rangle$
of $\rho$ with respect to $|\psi\rangle\langle\psi|$.
It is given by
\begin{eqnarray}
	F(\lambda)&=&
	{\rm tr}_{\tilde{\cal H}}\sum_i \bigl(
	\lambda\,\langle\psi|
	\left[(A_i\otimes B_i) (
	|\psi\rangle\langle\psi|
	\otimes\omega)
	(A_i\otimes B_i)^\dagger
	\right] |\psi\rangle
	\nonumber\\
	&+&
	(1-\lambda)\;\langle\psi|\left[
	(A_i\otimes B_i) (\zeta\otimes\omega)
	(A_i\otimes B_i)^\dagger\right] |\psi\rangle\bigr)/{\cal N},\nonumber
\end{eqnarray}
where
\begin{eqnarray}
	{\cal N}=\sum_i {\rm tr}\left[
	(A_i\otimes B_i)
	((\lambda|\psi\rangle\langle\psi|
	+(1-\lambda)
	\zeta)	
	\otimes\omega)
	(A_i\otimes B_i)^\dagger
	\right].\nonumber
\end{eqnarray}
%
$dF^2(\lambda)/d^2 \lambda = {\cal C}/{\cal N}^3$ with
a number $ {\cal C}$ independent of $\lambda$, and
one can argue as in the case of local operations without
a catalyst state \cite{Kent}:
The sign of the second derivative of the
function $f(\lambda)=
F(\lambda)-\lambda$ is constant for all $\lambda\in(0,1)$,
and therefore, this function is convex, concave or
linear. At $\lambda=0$, $f(0)\geq0$ as
$f(\lambda)\geq-\lambda$ for $\lambda\in(0,1)$, and
$f(1)\leq0$. $f(\lambda_0)\leq0$ follows from the
fact that the map Eq.\  (\ref{trans}) cannot transform
the state pertaining to $\lambda_0$
to an entangled state. Hence, $f(\lambda)\leq0$ for
all $\lambda\in[\lambda_0,1)$, i.e., the proportion
of $|\psi\rangle\langle\psi|$ can only decrease.

\section{Small transformations and catalysis for pure and mixed states}
%
So far, the findings in the pure state case and those
for mixed states have suggested a rather similar behavior of
both sets of states with respect to 
LOCC and ELOCC operations. However, things are
quite different in the next issue
concerning the possibility to enhance the 
range of accessible states with catalyst states
in ``small'' transformations.\\

\noindent {\bf Lemma 2. --} For all pure states $|\psi\rangle\in{\cal H}$
and all pure catalyst states $|\tilde\psi\rangle\in\tilde{\cal H}$
there exists a $\delta>0$ such that 
\begin{eqnarray}
\nonumber
|\psi\rangle\not\rightarrow |\phi\rangle{\rm \,\,under\,\,
LOCC}
\,\,\Rightarrow\,
|\psi\rangle\not\rightarrow |\phi\rangle{\rm \,\,under\,\,
ELOCC}
\nonumber
\end{eqnarray}
for all $|\phi\rangle\in {\cal H}$
with $|\langle\psi|\phi\rangle|^2>1-\delta$. 

{\it Proof:} Let $\alpha_1$, ..., $\alpha_N$
be the ordered lists of eigenvalues of 
${\rm tr}_A[|\psi\rangle\langle\psi|]$,
$N=\dim[{\cal H}_A]=\dim[{\cal H}_B]$, 
and let $\gamma_1$, ..., $\gamma_M$
be the corresponding list of the pure catalyst state,
$M=\dim[\tilde{\cal H}_A]=\dim[\tilde{\cal H}_B]$. 
Let now $\varepsilon>0$ 
and call an $\varepsilon$-list a list
$\beta_1$, ..., $\beta_N$ with
$1\geq \beta_1\geq ...\geq \beta_N\geq0$ that has the
property $|\beta_i-\alpha_i|<\varepsilon$ for all $i=1,...,N$.
There exists an $\varepsilon>0$ such that
for all $\varepsilon$-lists
$\beta_1$, ..., $\beta_N$
the statement that 
$\alpha_i\gamma_j>\alpha_k\gamma_l$ for some
$i,k\in\{1,...,N\}, j,l\in\{1,...,M\}$ implies that
$\beta_i\gamma_j>\beta_k\gamma_l$. This $\varepsilon$
is in the following referred to as $\tilde\varepsilon$.
Moreover, 
there exists a $\delta>0$ such that 
for each $|\phi\rangle\in {\cal H}$
with $|\langle\psi|\phi\rangle|^2>1-\delta$
the ordered
eigenvalues of ${\rm tr}_A[|\phi\rangle\langle\phi|]$
form a $\tilde\varepsilon$-list (and
hence, for such states it is not possible that
$\beta_i\gamma_j<\beta_k\gamma_l$ and
$\alpha_i\gamma_j>\alpha_k\gamma_l$). 
It follows that for all
such $|\phi\rangle\in {\cal H}$
with $|\langle\psi|\phi\rangle|^2>1-\delta$ 
the majorization relation
$
	{\rm 
	tr}_A[|\psi\rangle\langle\psi|\otimes|\tilde\psi\rangle\langle\tilde\psi|]
	\not\prec
{\rm tr}_A[|\phi\rangle\langle\phi|\otimes
|\tilde\psi\rangle\langle\tilde\psi|]
$ holds 
if ${\rm tr}_A[|\psi\rangle\langle\psi|]\not\prec
{\rm tr}_A[|\phi\rangle\langle\phi|]$. Finally, this implies
the statement of Lemma 2.
$\hfill{\Box}$

This is not true for mixed states, when the fidelity
of two states $\sigma$ and $\rho$ is
taken to be
$F(\sigma,\rho)= ({\rm tr} [(\sqrt{\sigma}\rho\sqrt{\sigma})^{1/2}])^2$
\cite{Uhlmann}.
Indeed,
there are states $\sigma\in{\cal S}({\cal H})$ such that
for every $\delta>0$ there are states $\rho\in{\cal S}({\cal H})$
with the property that $F(\sigma,\rho)>1-\delta$ and
$\sigma\not\rightarrow\rho$ under LOCC, but
$\sigma\rightarrow\rho$ under ELOCC. Such states can,
e.g., be constructed using the class of states
defined in Eq.\ (\ref{example}), Eq.\ (\ref{psi}), and
Eq.\ (\ref{phi}). For any  given $\delta>0$
there is a sufficiently small $\lambda>0$ such that 
the fidelity satisfies 
$F(\sigma,\rho)>1-\delta$.

Hence, quite surprisingly, in the case of 
entanglement manipulations
from an initial pure state to a close pure state
entanglement-assisted operations do not add any power to LOCC 
operations. To put it in different words, there
is no catalysis for sufficiently close pure states. 
Yet, for mixed states there can be catalysis for 
such close states.


\section{Conclusion and open problems}
In this paper we have 
investigated the
power of entanglement-assisted manipulation of
entangled quantum systems in mixed states. 
Interestingly, the counterintuitive class of ELOCC 
operations has proven to be superior to mere LOCC operations
also in the interior of the state space, for
which such strong tools as the majorization criterion
are not available. 
Yet, albeit these findings might contribute to the quest for
a better understanding of mixed-state entanglement, 
there are numerous open problems.
Stronger criteria for the possibility of certain entanglement
transformation are urgently needed. 
Finally, it is the hope that this
work will help to explore practical applications \cite{Barnum}
of the strange phenomenon of catalysis.

\section{Acknowledgements}
We would like to thank Martin B.\ Plenio and Julia Kempe
for very helpful hints.
We also acknowledge fruitful discussions with 
Daniel Jonathan and the participants of the A2 Consortial 
Meeting. This work was supported
by the European Union and the DFG.

\end{multicols}


\begin{thebibliography}{99}
\bibitem{HilbertRemark}
	The catalyst state is a density
	operator on the Hilbert space $\tilde {\cal H}=
	\tilde {\cal H}_A\otimes \tilde {\cal H}_B$, where
	$\tilde {\cal H}_A,\tilde {\cal H}_B =
	{\text{span}}\{|6\rangle,|7\rangle\}$.
\bibitem{MachineRemark}
It is further assumed that the purification
protocol is independent of
the parameter $\lambda$.
\end{thebibliography}
\end{document}